\documentstyle[prl,aps]{revtex}
\def\be{\begin{equation}} 
\def\ee{\end{equation}}
\def\ba{\begin{array}} 
\def\ea{\end{array}} 
\def\bea{\begin{eqnarray}}
\def\eea{\end{eqnarray}}

\begin{document}
\title{ SU(4) Chiral Quark Model with Configuration Mixing.}
\author{Harleen Dahiya  and Manmohan Gupta}
\address {Department of Physics, Centre of Advanced Study in Physics,
Panjab University, Chandigarh-160 014, India.} 
\date{\today}
 \maketitle

\begin{abstract}
Chiral quark model with configuration mixing and broken SU(3)$\times$U(1) 
symmetry has been extended to include the contribution from $c \bar c$
fluctuations by considering  broken SU(4) instead of SU(3). 
The implications of such a model have been studied for quark flavor and 
spin distribution functions corresponding to E866 and the NMC data. 
The predicted parameters regarding the charm spin distribution functions, 
for example,
$\Delta c$, $\frac{\Delta c}{{\Delta \Sigma}}$,  
 $\frac{\Delta c}{c}$
as well as the charm quark distribution functions, for example,
$\bar c$, $\frac{2\bar c}{(\bar u+\bar d)}$, $\frac{2 \bar
c}{(u+d)}$ and $\frac{(c+ \bar c)}{\sum (q+\bar q)}$ are  in 
agreement with other similar calculations. 
Specifically, we find $\Delta c=-0.009$, 
$\frac{\Delta c}{{\Delta \Sigma}}=-0.02$, $\bar c=0.03$ and 
$\frac{(c+ \bar c)}{\sum (q+\bar q)}=0.02$ for the $\chi$QM parameters
$a=0.1$, $\alpha=0.4$, $\beta=0.7$, $\zeta_{E866}=-1-2 \beta$, 
$\zeta_{NMC}=-2-2 \beta$ and $\gamma=0.3$, the latter appears due to the
extension of SU(3) to SU(4).
\end{abstract}

There has been considerable interest in estimating the possible size
of the intrinsic charm content of the nucleon
\cite{{charm1},{charm2},{charm3},{charm4},{charm5}}. 
Detailed investigations have been carried out regarding the size 
and implications of intrinsic charm contribution for nucleon 
\cite{songcharm}  in a version of chiral quark model ($\chi$QM) 
\cite{{wein},{manohar},{eichten},{cheng},{chengsu3},{song},{johan}}
which is quite successful in giving a satisfactory explanation of
``proton spin crisis'' {\cite{EMC}} including the violation of Gottfried 
sum rule {\cite{{GSR},{NMC},{E866}}}. Further, the same model
is also able to account for the existence of significant strange
quark content $\bar s$ \cite{{st q},{ao}} in the nucleon as well as is
able to provide fairly satisfactory explanation for 
quark flavor and spin distribution functions \cite{{adams}}, 
baryon magnetic moments {\cite{{manohar},{eichten},{cheng1}}}, 
absence of polarizations of the antiquark sea  in the nucleon 
{\cite{antiquark}},
hyperon decay parameters\cite{{close},{hyp1},{hyp2}} etc..

Recently, it has been shown that configuration mixing generated by
spin-spin forces \cite{{DGG},{Isgur},{yaouanc}}, 
known to be compatible with the $\chi$QM
\cite{{riska},{chengspin},{prl}}, improves the predictions of 
$\chi$QM regarding the quark distribution functions and the
spin polarization functions \cite{hd}. Further, $\chi$QM with 
configuration mixing (henceforth to be referred as $\chi$QM$_{gcm}$ ) 
when coupled with the quark sea polarization and orbital angular
momentum (Cheng-Li mechanism \cite{{cheng1}}) as well as
``confinement effects'' \cite{{effm1},{effm2}} 
is able to give an excellent fit \cite{hdorbit} to the octet magnetic 
moments and 
a perfect fit for the violation of Coleman Glashow sum rule \cite{cg}.

The successes of $\chi$QM in resolving the ``proton
spin crisis'' and related issues strongly suggest that
constituent quarks and the weakly interacting Goldstone bosons (GBs) 
provide the appropriate degrees of freedom 
in the nonperturbative regime of QCD. Thus
the quantum fluctuations generated by broken chiral symmetry in
$\chi$QM$_{gcm}$ should be able to provide a viable estimate of
the heavier quark flavor, for example,
$c \bar c$, $b \bar b$ and $t \bar t$. However, it is known that these
flavor fluctuations are much suppressed in the case of
$b \bar b$ and $t \bar t$ as compared to the $c \bar c$ 
because the intrinsic heavy quark contributions scale as $1/M^2_q$, where
$M_q$ is the mass of the heavy quark \cite{{charm1},{chengspin}}.
Therefore, regarding the intrinsic charm flavor content of the
nucleon one should estimate only the contribution of $c \bar c$ fluctuations 
and for that one should be considering the extension of SU(3) symmetry
in $\chi$QM to SU(4).

The purpose  of the  present  communication, on the one hand, is to 
extend $\chi$QM$_{gcm}$ with broken SU(3)$\times$U(1) symmetry to broken 
SU(4)$\times$U(1) symmetry. On the other hand, using the NMC \cite{NMC}
and the latest E866 data \cite{E866},  we intend to study the 
implications of such a model for quark flavor and spin distribution 
functions, in particular the charm quark flavor and spin distribution 
functions.

The details of $\chi$QM$_{gcm}$ within the SU(3) framework have already
been discussed in Ref. \cite{hd}, here we discuss the essentials of its 
extension to SU(4) $\chi$QM$_{gcm}$.   
To begin with, the basic process in the $\chi$QM is the emission of a  
GB which further  splits into $q \bar q$ pair, 
for example,
\be
q_{\pm} \rightarrow {\rm GB}^{0} + q^{'}_{\mp}  \rightarrow (q \bar q^{'})   
+q_{\mp}^{'}, \label{basic}
\ee
wherein $q \bar q^{'}$ pairs and  $q^{'}$ constitute the ``quark sea'' with
$q^{'}$ having opposite helicity as that of $q$.
The effective   Lagrangian  describing
interaction between quarks and the mesons in the SU(4) case is

\be  
{\cal L}  =  g_{15} \bar  q \phi  q,  \label{lag}
\ee 
where  $g_{15}$ is  the
coupling constant,
\[ q =\left( \ba{c} u \\ d \\ s \\c \ea \right)\]
and $\phi$ represents the SU(4) matrix
\[
\phi=\left( \ba{cccc} \frac{\pi^o}{\sqrt 2}
+\beta\frac{\eta}{\sqrt 6}+\zeta\frac{\eta^{'}}{4\sqrt 3} 
-\gamma\frac{3 {\eta}^o_c}{4} & \pi^+ & \alpha K^+ & \gamma \bar D^o  \\  
\pi^- &  -\frac{\pi^o}{\sqrt  2}   +\beta  \frac{\eta}{\sqrt  6}
+\zeta\frac{\eta^{'}}{4\sqrt 3}-\gamma\frac{3  {\eta}^o_c}{4} & \alpha
K^o  &  \gamma  D^-             \\  
\alpha  K^-  &  \alpha \bar{K}^o  &  -\beta
\frac{2\eta}{\sqrt  6}  +\zeta\frac{\eta^{'}}{4\sqrt  3}-\gamma\frac{3
{\eta}^o_c}{4} & \gamma D^-_s            \\  
\gamma D^o &\gamma D^+ &\gamma D^+_s
& -\zeta\frac{3\eta^{'}}{4\sqrt 3}+\gamma\frac{3{\eta}^o_c}{4}  
\ea \right). \]

SU(4) symmetry  breaking is introduced by  considering different quark
masses $M_c >  M_s > M_{u,d}$ as well as by  considering the masses of
GBs to be non-degenerate $(M_{D} > M_{K,\eta} > M_{\pi})$
similar to the SU(3) case {\cite{{chengsu3},{song},{johan},{cheng1}}}, 
whereas  the axial U(1)  breaking is
introduced by $M_{\eta^{'}} > M_{K,\eta}$
{\cite{{cheng},{song},{johan},{cheng1}}}.         The        parameter
$a(=|g_{15}|^2$)   denotes  the   transition  probability   of  chiral
fluctuation of  the splittings  $u(d) \rightarrow d(u)  + \pi^{+(-)}$,
whereas $\alpha^2 a$, $\beta^2 a$, $\zeta^2 a$ and $\gamma^2 a$ denote
the  probabilities of  transition of  $u(d) \rightarrow  s  + K^{-(o)}$,
$u(d,s)  \rightarrow  u(d,s) +  \eta$,  $u(d,s)  \rightarrow u(d,s)  +
\eta^{'}$ and $u(d) \rightarrow c + \bar D^o(D^-)$ respectively.

The detailed effects of configuration mixing generated by spin-spin forces
\cite{{DGG},{Isgur},{yaouanc}} in the context of $\chi$QM has already been 
discussed in $\chi$QM$_{gcm}$ \cite{hd}, however to make the manuscript 
readable as well as self contained
we include here some of the essentials of configuration mixing.
Following reference \cite{hd}, the wavefunction for the octet of baryons 
after configuration mixing is given as
\be
|B \rangle=\left(|56,0^+\rangle_{N=0} \cos \theta +|56,0^+ \rangle_{N=2}  
\sin \theta \right) \cos \phi 
+  \left(|70,0^+\rangle_{N=2} \cos \theta +|70,2^+\rangle_{N=2}  
\sin \theta \right) \sin \phi, \label{full mixing}
\ee
where $\theta$ and $\phi$ are the mixing angles with
\bea
 |56,0^+\rangle_{N=0,2} &=& \frac{1}{\sqrt 2}(\chi^{'} \phi^{'} +
\chi^{''} \phi^{''}) \psi^{s}(0^+), \label{56}   \\      
|70,0^+\rangle_{N=2} &=&  \frac{1}{2}[(\phi^{'} \chi^{''} +\phi^{''}\chi^{'})
\psi^{'}(0^+) + (\phi^{'} \chi^{'} -\phi^{''} \chi^{''})\psi^{''}(0^+)], 
\label{70}  \\ 
|70,2^+\rangle_{N=2} &=&  \frac{1}{\sqrt 2}[ \phi^{'} \chi^{s} \psi^{'}(2^+)
+ \phi^{''} \chi^{s} \psi^{''}(2^+)].                                    
\eea
The spin wave functions  are as follows

\[ \chi^{'} =  \frac{1}{\sqrt 2}(\uparrow \downarrow \uparrow
-\downarrow \uparrow \uparrow),~~~
\chi^{''} =  \frac{1}{\sqrt 6} (2\uparrow \uparrow \downarrow
-\uparrow \downarrow \uparrow -\downarrow \uparrow \uparrow). \] \\
The isospin wave functions for proton are
\[\phi^{'} = \frac{1}{\sqrt 2}(udu-duu),~~~
\phi^{''} = \frac{1}{\sqrt 6}(2uud-udu-duu).\]
The above mixing can
effectively be reduced to non trivial mixing 
{\cite{{yaouanc},{effm2},{mgupta1}}} and the corresponding
``mixed'' octet of baryons is expressed as

\begin{equation}
|B\rangle 
\equiv \left|8,{\frac{1}{2}}^+ \right\rangle 
= \cos \phi |56,0^+\rangle_{N=0}
+ \sin \phi|70,0^+\rangle_{N=2}.  \label{mixed}
\end{equation} 
Henceforth, we would not distinguish between configuration mixing 
given in Eq. (\ref{full mixing}) and the ``mixed'' octet given above. 

Following Ref. \cite{{chengsu3},{johan}}, the total probability of 
no emission of GB from a $q$ quark ($q=u,~d,~ s,~c$) 
can be calculated from the Lagrangian for the SU(4) case and is given by
\be
P_q=1-\sum P_q, \label{probability}
\ee
where
\bea
\sum P_u &=& a \left
(\frac{3}{2}+\alpha^2+\frac{\beta^2}{6}+\frac{\zeta^2}{48}+\gamma^2
\right ), \\
\sum P_d &=& a \left
(\frac{3}{2}+\alpha^2+\frac{\beta^2}{6}+\frac{\zeta^2}{48}+\gamma^2
\right ), \\
\sum P_s &=& a \left ( 2 \alpha^2+\frac{2}{3}
\beta^2+\frac{\zeta^2}{48}+\gamma^2 \right ), \\
\sum P_c &=& a \left
(\frac{3}{16} \zeta^2+\frac{57}{16} \gamma^2 \right ).
\eea

Before getting into the details of the calculations one needs to formulate
experimentally measurable quantities having implications for the charm 
content of the nucleon as well as dependent on the unpolarized
quark distribution functions and the  spin polarization functions in 
$\chi$QM$_{gcm}$ with broken SU(4) symmetry. 
We first
calculate the spin polarizations and the related quantities which
are affected by the ``mixed'' nucleon.
The spin structure of a nucleon for the SU(4) case is defined in a 
similar manner as that of the SU(3) case \cite{{cheng},{song},{johan}}
and is 
\be
\hat B \equiv \langle B|N|B\rangle,
\ee
where $|B\rangle$ is the nucleon wavefunction defined in Eq. (\ref{mixed})
and $N$ is the number operator given by
\be
 N=n_{u^{+}}u^{+} + n_{u^{-}}u^{-} +
n_{d^{+}}d^{+} + n_{d^{-}}d^{-} +
n_{s^{+}}s^{+} + n_{s^{-}}s^{-}+
n_{c^{+}}c^{+} + n_{c^{-}}c^{-}, 
\ee
where $n_{q^{\pm}}$ are the number of 
$q^{\pm}$ quarks. 
The spin structure of the ``mixed'' nucleon, defined through the
 Eq. (\ref{mixed}), is given by
\be
 \left\langle 8,{\frac{1}{2}}^+|N|8,{\frac{1}{2}}^+\right\rangle=\cos^2 \phi
\langle56,0^+|N|56,0^+\rangle+\sin^2 \phi\langle70,0^+|N|70,0^+\rangle. \label{spinst}
\ee
Using Eqs. (\ref{56}) and (\ref{70}), for proton we get
\bea
 \langle 56,0^+|N|56,0^+\rangle&=&\frac{5}{3} u^{+} +\frac{1}{3} u^{-}+
\frac{1}{3} d^{+} +\frac{2}{3} d^{-}, \label{56st} \\ 
\langle 70,0^+|N|70,0^+\rangle& =& \frac{4}{3} u^{+} +\frac{2}{3} u^{-}+
\frac{2}{3} d^{+} +\frac{1}{3} d^{-}. \label{70st}
\eea

The spin structure
after one interaction can be obtained by  substituting in the above 
equations for every quark, for example,

\be
 q^{\pm} \rightarrow P_q q^{\pm} + |\psi(q^{\pm})|^2, \label{sea st}
\ee
where $P_q$ is the probability of no emission of GB from a $q$ quark
defined in Eq. (\ref{probability}) and the 
probabilities of transforming a $q^{\pm}$ quark are
$|\psi(q^{\pm})|^2$ which are given for the SU(4) broken $\chi$QM as

\bea
 |\psi(u^{\pm})|^2&=&a \left ( \frac{1}{2}+\frac{\beta^2}{6}
+\frac{\zeta^2}{48}+\frac{\gamma^2}{16} \right ) u^{\mp}+    
a d^{\mp}+a \alpha^2 s^{\mp} +a \gamma^2
c^{\mp},\\ 
 |\psi(d^{\pm})|^2 &=&a u^{\mp}+ a \left (
\frac{1}{2}+\frac{\beta^2}{6} +\frac{\zeta^2}{48}+\frac{\gamma^2}{16}
\right ) d^{\mp}+ a \alpha^2 s^{\mp}+a \gamma^2
c^{\mp},\\   
 |\psi(s^{\pm})|^2&=&   a \alpha^2 u^{\mp}+ 
a \alpha^2 d^{\mp}+ a \left ( \frac{2}{3}\beta^2
 +\frac{\zeta^2}{48}+\frac{\gamma^2}{16} \right )s^{\mp}+a \gamma^2
c^{\mp}, \\  
 |\psi(c^{\pm})|^2&=&   a \gamma^2 u^{\mp}+ 
a \gamma^2 d^{\mp}+a \gamma^2 s^{\mp}+ 
a \left (\frac{3}{16} \zeta^2 + \frac{9}{16}\gamma^2 \right
)c^{\mp}.
\eea
Substituting Eqs. (\ref{56st}), (\ref{70st}) and (\ref{sea st})  
in Eq. (\ref{spinst}), we can derive
the spin polarizations, defined as 
$ \Delta q=q^{+}-q^{-}+{\bar q}^+-{\bar q}^-$, for example
\bea 
\Delta u&=&  \cos^2   \phi  \left[\frac{4}{3}-
\frac{a}{3}  (7+4  \alpha^2+ \frac{4}{3} \beta^2  + 
\frac{1}{6} \zeta^2+\frac{9}{2}\gamma^2)\right]  
+  \sin^2 \phi \left[\frac{2}{3}-\frac{a}{3} (5+2
\alpha^2+ \frac{2}{3} \beta^2 + \frac{1}{12}
\zeta^2+\frac{9}{4}\gamma^2)\right], \label{delta u} \\  
\Delta d &=&\cos^2 \phi
\left[-\frac{1}{3}-\frac{a}{3}     (2-\alpha^2-    \frac{1}{3}\beta^2-
\frac{1}{24} \zeta^2-\frac{9}{8}\gamma^2)\right] + 
\sin^2  \phi  \left[\frac{1}{3}-\frac{a}{3}  (4+\alpha^2+  \frac{1}{3}
\beta^2 +  \frac{1}{24} \zeta^2+\frac{9}{8}\gamma^2)\right], 
\label{delta d} \\ 
\Delta s &=& -a \alpha^2, \\ 
\Delta c &=& -a \gamma^2.  
\eea

After  having formulated the spin  polarizations of various  quarks in
terms of SU(4) $\chi$QM$_{gcm}$, we consider several measured quantities 
which are expressed in terms of the above mentioned spin  polarization 
functions.
Some of the quantities usually calculated in the $\chi$QM are the
weak axial-vector form factors and are expressed as
\bea
(G_A/G_V)_{n \rightarrow p}=\Delta_3 &=& \Delta u-\Delta d,\\
(G_A/G_V)_{\Lambda \rightarrow p} &=& \frac{1}{2}
(2 \Delta u-\Delta d-\Delta s), \\
(G_A/G_V)_{\Sigma^- \rightarrow n} &=& \Delta d-\Delta s, \\
(G_A/G_V)_{\Xi^- \rightarrow \Lambda} &=& \frac{1}{3}(\Delta u+\Delta d-2 \Delta s).
\eea
Another quantity which is usually evaluated is the total helicity 
fraction carried by the quark $q$  defined as 
\be 
\Delta q/{\Delta \Sigma},
\ee
where
\be 
\Delta \Sigma= \Delta_0= \Delta u+\Delta  d+\Delta s+\Delta c,
\ee 
which is the total quark spin content. It may be added that the 
expressions for Bjorken \cite{bjorken} and
 Ellis-Jaffe sum rules \cite{ellis} are not affected in the present case,
however, the contributions to these get affected as $\Delta u$ and 
$\Delta d$ include contributions from charm quark fluctuations also. 
For the sake of completeness we express these  in
terms of the above mentioned
spin polarization functions, for example, the Bjorken sum rule is
\be
\int_0^1[g_1^p(x,Q^2)-g_1^n(x,Q^2)]dx=\frac{G_A/G_V}{6},
\ee
where $g_1^{p(n)}(x,Q^2)$ is the spin structure function of the proton 
(neutron) and $G_A/G_V$ is the $\beta$ decay constant for neutron.
The  Ellis-Jaffe sum rule is given by
\be 
\Delta_8=\Delta u+\Delta d-2\Delta s=3F-D,
\ee
where $F$ and $D$ are the axial coupling constants estimated from the 
weak decays of hyperons and their relation to
the spin polarization functions remain the same in the case of 
SU(3) and SU(4) $\chi$QM, for example,
\bea 
F&=& \frac{1}{2}(\Delta u-\Delta  s), \\ 
D&=&\frac{1}{2}(\Delta u-2\Delta d+\Delta s).  
\eea
However, again $\Delta u$ and $\Delta d$ take on different values compared 
to SU(3) because there is an additional term corresponding to the charm quark 
fluctuations and also the coefficient corresponding to the $\eta'$ term 
is different. This can be seen by  comparing the expressions for 
$\Delta u$ and $\Delta d$ in the present case
(Eqs (\ref{delta u}) and (\ref{delta d})) and the corresponding expressions
for $\Delta u$ and $\Delta d$ in the SU(3) case  \cite{hd}.

The  unpolarized valence quark distribution functions are not affected 
by configuration mixing, however these get affected due to the addition 
of $c \bar c$ fluctuations and hence are dependent on the SU(4)$\times$U(1)
symmetry breaking parameters.
A calculation of these quantities also assumes importance in the present 
case as we attempt to effect a unified fit to spin and quark 
distribution functions. The quark distribution functions which have
implications for the symmetry breaking parameters of SU(4) are the
antiquark flavor contents of the ``quark sea'' which
can be expressed as  \cite{{songcharm},{cheng},{johan}} 
\bea 
\bar u   &=&\frac{1}{48}[(2 \beta+ \zeta+2)^2 +80]   a ,  \label{bar u} \\  
\bar d &=&\frac{1}{48}[(2 \beta +\zeta -2)^2+128]   a , \label{bar d} \\   
\bar s &=&\frac{1}{48}[(\zeta-4 \beta)^2 +144 {\alpha}^{2}] a ,
\label{bar s}  \\  
\bar c &=&\frac{51}{16}\gamma^2 a . \label{bar c}  
\eea  
The deviation from the Gottfried  sum rule \cite{{GSR},{NMC},{E866}} 
 can be expressed in terms of the 
symmetry breaking parameters $\beta$  and $\zeta$ as
\be
\left[I_G-\frac{1}{3}\right]=
\frac{2}{3}\left[\frac{a}{6}( 2\beta+ \zeta-6)\right], \label{zeta}
\ee
where $I_G=\int_{0}^{1}dx\frac{F_2^p(x)-F_2^n(x)}{x}$ is the Gottfried 
integral.
Similarly, $\bar d/\bar u$ {\cite{{E866},{na51}}} measured 
through  the ratio of  muon pair production
cross sections  $\sigma_{pn}$ and $\sigma_{pp}$, is expressed in the 
present case as follows 
\be
\bar d/\bar u=\frac{(2 \beta +\zeta -2)^2+128}{(2 \beta+ \zeta+2)^2 +80}.
\ee
Further, some of the quark flavor fractions usually discussed in the 
literature \cite{{charm2},{charm4},{charm5},{ao}} 
are also expressed as follows
\bea 
f_q&=&\frac{q+\bar  q}{[\sum_{q} (q+\bar  q)]}, \\
\frac{2 \bar s}{\bar u+ \bar d}&=&\frac{(\zeta-4 \beta)^2 +144 \alpha^2}
{(2\beta+\zeta)^2+108}, \\
\frac{2\bar c}{\bar u+ \bar d}&=&\frac{153\gamma^2}{(2\beta+\zeta)^2+108}, \\
\frac{2 \bar s}{u+d} &=&\frac{a[(\zeta-4 \beta)^2 +144 \alpha^2]}
{72+a[(2\beta+\zeta)^2+108]}, \\
\frac{2 \bar c}{u+d}&=& \frac{153 \gamma^2 a}{72+a[(2\beta+\zeta)^2+108]}.
\eea
The $q$ and $\bar q$ distributions in the case of proton  are normalized 
as
\be 
u-\bar u=2, ~~~d-\bar d=1,  ~~~ s-\bar s=0,~~~c-\bar c=0. 
\ee 

The above mentioned spin polarization functions and the quark distribution 
functions are to be fitted for the E866 as well as NMC data. In principle, 
one can obtain this fit by considering all possible variations of SU(4) and  
U(1) symmetry breaking parameters as well as the mixing angle $\phi$. However, 
keeping in mind the general expectation that the $c \bar c$ contribution 
cannot be large compared to the $s \bar s$ contribution  
as well as to compare our results with the corresponding results of 
$\chi$QM$_{gcm}$ with SU(3)$\times$U(1) symmetry breaking, 
in our analysis we have considered the parameters $a$, $\alpha$ and $\beta$
to be the same as in SU(3) case.  The parameter $\gamma$, controlling 
$c \bar c$ contribution, has been varied from 0.1 to 0.3 as
considered by other authors \cite{songcharm}.
The parameter $\zeta$, as discussed in Ref. \cite{{cheng},{johan}}, 
represents the U(1) symmetry breaking parameter and is
responsible for reproducing the violation of  Gottfried sum rule. 
Similar to the case of SU(3), here also we have derived the relation of 
$\zeta$ in terms of $\beta$ from the violation 
of  Gottfried sum rule given in Eq. (\ref{zeta}). Its value undergoes 
a major change in case of SU(4) as compared to that of
SU(3), for example, for E866 we have
$\zeta= -1-2\beta$ as compared to $\zeta=-0.3-\beta/2$ and
for NMC we have $\zeta= -2-2\beta$ as compared to $\zeta=-0.7-\beta/2$.
The mixing parameter $\phi$ is taken to be 20$^o$ as found from
neutron charge radius \cite{{yaouanc},{neu charge}}  and considered in our earlier work 
\cite{{hd},{hdorbit}}.
The results without configuration mixing
can easily be obtained by substituting $\phi=0$ in the expressions for
spin polarization functions.

In Tables (\ref{spin}) and (\ref{quark}) we have presented the results 
of our calculations
pertaining to spin distribution and quark distribution functions respectively.
In the tables we have also included the results  of $\chi$QM$_{gcm}$
with SU(3) symmetry breaking, primarily to compare these with the 
corresponding SU(4) symmetry breaking calculations. 
Similarly, in the case of SU(4) we have also 
included the results without configuration mixing which allows the 
comparison of the two results for the SU(4) case. The
calculations have been performed for E866 as well as for NMC data 
and the corresponding results for each case have been included in 
the tables.

From Table (\ref{spin}), one can immediately find out that
calculations in SU(4) $\chi$QM$_{gcm}$ are able to maintain the agreement 
achieved in the case of SU(3) $\chi$QM$_{gcm}$ for
the spin polarizations $\Delta u$, $\Delta d$,  $\Delta s$ and the related 
quantities such as $\Delta \Sigma/2$, $\Delta_8$, hyperon decay parameters F 
and D as well as the weak axial-vector form factors. 
Expectedly, the two results look to be similar with 
the slight changes occurring in the case
of SU(4) in comparison to the SU(3) results primarily due to
changes induced in $\Delta u$ and $\Delta d$ involving the symmetry
breaking parameters, $\zeta$ and $\gamma$, which take different values
as compared to SU(3) $\chi$QM.
One also finds that the quantities involving strange quarks are not
affected because in the present formalism there is no 
process which can mix the strange and charm contributions.
The main predictions of SU(4) $\chi$QM pertains to quantities such as, 
$\Delta c$, $\Delta c/\Delta \Sigma$ and $\Delta c/c$, which have
vanishing amplitudes at the tree level in SU(3) $\chi$QM.
The charm contribution, corresponding to spin polarization
functions, however is smaller by one order of magnitude as
compared to the corresponding parameter involving strange quark which is 
in accord with another recent analysis \cite{songcharm}.
Similarly, we find that 
$\Delta_{15} (=\Delta u+\Delta d+\Delta s-3 \Delta c)$ 
is also in agreement with analysis of Ref. \cite{songcharm}. 

The role of configuration mixing in the present context can easily be
examined from Table (\ref{spin}). We find that the configuration mixing 
effects a uniform improvement in the case of spin polarization functions
compared to those without configuration mixing. For example, 
$\Delta u$, $\Delta_8$, hyperon decay parameters $F$ and $D$,
weak axial-vector form factors $(G_A/G_V)_{n
\rightarrow p}$, $(G_A/G_V)_{\Lambda \rightarrow p}$ show
remarkable improvement whereas the results of
$\Delta s$, $\Delta c$, $\Delta \Sigma$,
$(G_A/G_V)_{\Sigma^- \rightarrow n}$ and 
$(G_A/G_V)_{\Xi^- \rightarrow \Lambda}$ are in good deal of 
agreement with the data.

From Table (\ref{quark}), we find that in the SU(4) $\chi$QM the
important measurable quark distribution functions, 
for example, $\bar d-\bar u$, $\bar d/\bar u$, $I_G$,
$\frac{2 \bar s}{(\bar u+\bar d)}$, $\frac{2 \bar s}{(u+d)}$, $f_s$,
$f_3/f_8$ etc.  are in good agreement
with the data. 
It may be noted that although in the present case some of the symmetry
breaking parameters take different values compared to SU(3) case still
we find that our results are in good agreement with the data. It may also
be noted that the quark distribution functions  $\bar u$, $\bar d$ and
$\bar s$ assume different values in the present case as compared to the 
SU(3) case because of the changed parameters,
however the quantities dependent on these again remain in good
agreement with data. The values of the quantities
involving $c$ quark, for example,  $\bar c$, 
$\frac{2\bar c}{(\bar u+\bar d)}$, $\frac{2 \bar
c}{(u+d)}$ and $f_c$, are in agreement with the predictions given by other
authors \cite{{charm2},{charm3},{charm4}}. 
Interestingly, in the case of $\frac{2 \bar s}{(\bar u+\bar d)}$ and 
$\frac{2 \bar s}{(u+d)}$, our predictions are in better agreement with 
data compared to another recent analysis \cite{songcharm}, 
therefore a refinement in the measurement of these would
have important implications for the details of the $\chi$QM.

A closer scrutiny of the tables reveals several additional points. There
is a difference in the predictions of the fit corresponding to E866 and
NMC data as is evident in the case of $\Delta u$, $\Delta_3$, 
 $(G_A/G_V)_{n \rightarrow p}$, $(G_A/G_V)_{\Lambda \rightarrow p}$,
$\bar s$, $\frac{2\bar s}{(\bar u+\bar d)}$, $\frac{2 \bar s}{(u+d)}$, 
$f_s$, $f_3$, $f_8$ and $\frac{\sum \bar q}{\sum q}$. 
This is primarily due to the fact that the parameter $\zeta$,
responsible for fitting the violation of
Gottfried sum rule, assumes different values in the two cases.
It may be of interest to mention that in SU(4) $\chi$QM$_{gcm}$, 
$\frac{\Delta c}{c}$ is independent of any splitting parameter unlike 
other similar fractions for different quark flavors
$\frac{\Delta q}{q}$.

To summarize, we have extended the  $\chi$QM$_{gcm}$ with broken
SU(3)$\times$U(1) symmetry to  broken SU(4)$\times$U(1) symmetry with
and without configuration mixing. 
The implications of such a model have been studied for quark flavor and 
spin distribution functions corresponding to NMC and the latest E866 data.
The charm dependent quantities such as charm spin distribution functions 
$\Delta c$, $\frac{\Delta c}{{\Delta \Sigma}}$,  
 $\frac{\Delta c}{c}$
and the charm quark distribution functions
$\bar c$, $\frac{2\bar c}{(\bar u+\bar d)}$, $\frac{2 \bar
c}{(u+d)}$ and $\frac{(c+ \bar c)}{\sum (q+\bar q)}$, have been calculated
and the results are  in agreement with other 
similar calculations.
Specifically, we find $\Delta c=-0.009$, 
$\frac{\Delta c}{{\Delta \Sigma}}=-0.02$, $\bar c=0.03$ and 
$\frac{(c+ \bar c)}{\sum (q+\bar q)}=0.02$.
Interestingly,  the SU(4) results remain in agreement with that corresponding 
to the SU(3) calculations despite different values of some of the symmetry 
breaking parameters. It may also be noted that the 
results with configuration mixing generally show better overlap with data than
those without configuration mixing.

In conclusion, we would like to mention that $\chi$QM with broken
SU(4) symmetry, apart from maintaining the successes of 
$\chi$QM with broken SU(3) symmetry, predicts the intrinsic charm spin
and flavor distribution content of the nucleon which is found to be almost an
order of magnitude smaller than the strange quark contributions but not 
entirely insignificant.
A measurement of these charm related quantities would not only test the 
$\chi$QM but would also provide an insight into the nonperturbative 
regime of QCD.

\vskip .2cm
  {\bf ACKNOWLEDGMENTS}\\
The authors would like to thank S.D. Sharma and M. Randhawa for a few 
useful discussions. 
H.D. would like to thank CSIR, Govt. of India, for
 financial support and the chairman,
 Department of Physics, for providing facilities to work
 in the department.

\pagebreak

\begin{table}
\begin{center}
\begin{tabular}{cccccccc}      
 &  &\multicolumn{2}{c} {SU(3)} &
\multicolumn{4}{c} {SU(4)}\\ \cline{3-8} 
Quantity & Data & 
\multicolumn{2}{c} {$\chi$QM$_{gcm}$} &\multicolumn{2}{c} {$\chi$QM}  &
\multicolumn{2}{c} {$\chi$QM$_{gcm}$} \\ \cline{3-8}

& & NMC &E866  & NMC & E866 & NMC & E866 \\ \hline
$\Delta u$ &0.85 $\pm$ 0.04 \cite{adams}&0.91 & 0.92 & 0.98 & 1.01
& 0.92 &0.94 \\
$\Delta d$ &$-$0.41 $\pm$ 0.04\cite{adams}&-0.33 & -0.34 & -0.37 &
-0.38 & -0.31 & -0.32 \\
$\Delta s$ &$-$0.07 $\pm$ 0.04\cite{adams}&-0.02 & -0.02 & -0.02 &
-0.02 & -0.02 & -0.02  \\
$\Delta c$ &$-0.3^*$ \cite{charm3}&&&&&& \\
           &$-0.02 \pm 0.004^*$\cite{charm3}& 0 & 0 & -0.009 & -0.009 &
-0.009 & -0.009 \\
           &$-5\times 10^{-4^*}$\cite{charm3}&&&&&& \\
$\Delta_0/2=\Delta \Sigma/2$ &0.19 $\pm$ 0.06\cite{adams}& 0.28 & 0.28 & 0.29 &
0.30 & 0.29 & 0.30  \\
$\Delta_3=G_A/G_V$ &1.267 $\pm$ 0.0035\cite{PDG}& 1.24 & 1.26 & 1.35 & 1.39 &
1.23 & 1.26 \\
$\Delta_8$ &0.58 $\pm$ 0.025\cite{PDG}&0.62 & 0.62 & 0.65 & 0.67 & 
0.65 &0.66 \\

$\Delta_{15}$ & & 0.56 & 0.56 & 0.62 & 0.64 & 
0.62 &0.63 \\
$\Delta u/{\Delta \Sigma}$ & &1.62 & 1.64 & 1.69 & 1.68 & 1.59 & 1.57  \\
$\Delta d/{\Delta \Sigma}$ && -0.59 & -0.61 &-0.64& -0.63 & -0.53 & -0.53 \\
$\Delta s/{\Delta \Sigma}$ && -0.03 & -0.03 & -0.03 & -0.03 & -0.03 &
-0.03 \\
$\Delta c/{\Delta \Sigma}$ &$-0.08 \pm 0.01^*$\cite{charm3}&0 & 0 &-0.02
&-0.02 & -0.02 & -0.02  \\
                           &$-0.033^*$\cite{charm3}&&&&&& \\
$\Delta u/u$ & & 0.41 & 0.42 & 0.45 & 0.46 & 0.42 & 0.43  \\
$\Delta d/d$ & &-0.25 & -0.26 & -0.28 & -0.29 & -0.24 & -0.25  \\
$\Delta s/s$ &&-0.14 & -0.20 & -0.16 & -0.19 & -0.16 & -0.19 \\
$\Delta c/c$ &&0 & 0 & -0.30 & -0.30 & -0.30 & -0.30  \\ \hline

$F$ &0.462 & 0.47 & 0.475   & 0.51 & 0.51  & 0.48  & 0.485 \\ 
$D$ &0.794 & 0.78  & .785  & 0.88  &  0.87 & 0.78   & 0.775  \\ 
$F/D$ &0.575 &  0.60 & 0.605  &  0.582 & 0.586  & 0.615 & 0.61  \\ \hline
$(G_A/G_V)_{n \rightarrow p}$ & 1.26 $\pm$ 0.0035 \cite{PDG}&
1.24 &1.26 & 1.35 &1.39 & 1.23 & 1.26 \\
$(G_A/G_V)_{\Lambda \rightarrow p}$ & 0.72 $\pm$ 0.02 \cite{PDG} & 0.72 & 
0.73 & 0.78 & 0.81 & 0.72 & 0.74 \\
$(G_A/G_V)_{\Sigma^- \rightarrow n}$ & -0.34 $\pm$ 0.02 \cite{PDG}& -0.31 &
-0.32 & -0.35 & -0.36 & -0.29 & -0.30 \\
$(G_A/G_V)_{\Xi^- \rightarrow \Lambda}$ & 0.25 $\pm$ 0.05 \cite{PDG} & 0.22 &
0.21 & 0.22 &0.22 &0.22 &0.22 \\

\end{tabular}
\end{center}
{\small * These values correspond to the calculated values.}
\caption{The calculated values of
spin polarization functions $\Delta u, ~\Delta d, ~\Delta s, ~\Delta c$,
quantities dependent  on these: $\Delta \Sigma$, $G_A/G_V$,
$\Delta_8$ and the hyperon decay parameters both for  NMC and E866 data. } 
\label{spin}
\end{table}

\begin{table}
\begin{center}
\begin{tabular}{cccccc}     
 &   &\multicolumn{2}{c} {SU(3) $\chi$QM} &
\multicolumn{2}{c} {SU(4) $\chi$QM}\\ \cline{3-6} 
Quantity & Data&  NMC  & E866  & NMC & E866 \\ \hline 
$\bar u$ && 0.183 & 0.189&0.167 & 0.169 \\
$\bar d$ && 0.33  & 0.31 &0.30 & 0.285  \\
$\bar s$ && 0.14 & 0.10 &0.128  & 0.104   \\
$\bar c$ &&0 & 0 & 0.03  & 0.03  \\
$\bar d-\bar u$ &.147 $\pm$ 0.039\cite{NMC}& 0.147 & 0.117&0.133 & 0.117 \\
		&.118 $\pm$ 0.018\cite{E866}&&&& \\
$\bar d/\bar u$ &1.96 $\pm$ 0.246\cite{na51}& 1.89 & 1.59 & 1.80 & 1.69 \\
		&1.41 $\pm$ 0.146\cite{E866}&&&& \\
$I_G$ &.235 $\pm$ 0.005\cite{NMC}& 0.235 & 0.255 & 0.244 & 0.256 \\
		&.259 $\pm$ 0.005\cite{E866}&&&& \\
$\frac{2 \bar s}{(\bar u+\bar d)}$ &.477 $\pm$ 0.051\cite{ao}& 0.55 & 0.41 
& 0.55 & 0.46  \\
$\frac{2 \bar c}{(\bar u+\bar d)}$ && 0 & 0 & 0.123  & 0.126  \\
$\frac{2 \bar s}{(u+d)}$ &.099 $\pm$ 0.009\cite{ao}& 0.08 & 0.06 & 
0.07  & 0.06  \\
$\frac{2 \bar c}{(u+d)}$ && 0 & 0 & 0.02 & 0.02 \\
$f_u= \frac{(u+ \bar u)}{\sum (q+\bar q)}$ &&0.65 & 0.66 & 0.64 & 0.65 \\
$f_d= \frac{(d+ \bar d)}{\sum (q+\bar q)}$ &&0.45 & 0.45 &0.44  & 0.44  \\
$f_s= \frac{(s+ \bar s)}{\sum (q+\bar q)}$ &.076 $\pm$0.02\cite{ao} & 
0.08 & 0.06 & 0.07  & 0.06    \\
$f_c= \frac{(c+ \bar c)}{\sum (q+\bar q)}$ &0.03$^*$ \cite{charm5}&&&& \\
 	&0.02$^*$\cite{charm2} &0 & 0 &0.02  & 0.02  \\
            &0.01$^*$ \cite{charm4}&&&& \\
$f_3=f_u-f_d$ & & 0.20 & 0.21 & 0.20  & 0.21 \\
$f_8=f_u+f_d-2 f_s$ & & 0.94  & 0.99  & 0.94  & 0.97 \\
$f_3/f_8$ & .21 $\pm$ .05 \cite{cheng}& 0.21 & 0.21  & 0.21  & 0.22 \\
$\frac{\sum \bar q}{\sum q}$ &0.245 $\pm$ 0.005 \cite{ao}&0.178 & 0.166 &
0.172 & 0.164 \\ 

\end{tabular}
\end{center}
\caption{The calculated values of quark flavor distribution functions 
and other dependent quantities as calculated in the SU(3) and SU(4) 
$\chi$QM with symmetry breaking with the 
same values of symmetry breaking parameters as used in spin distribution 
functions and hyperon $\beta$ decay parameters.} \label{quark}
\end{table}

\end{document}